\documentclass[prl,aps,twocolumn,amsmath,amssymb]{revtex4}
\usepackage{graphicx}
\usepackage{dcolumn}
\usepackage{bm}
\usepackage{times}
\usepackage{epstopdf}
\usepackage{amsmath}
\usepackage{epsfig}
\usepackage{comment}
\newcommand{\ket}[1]{\left\vert#1\right\rangle}

\newcommand{\eq}{Eq.~}

\newcommand{\eqs}{Eqs.~}
\newcommand{\fig}{Fig.~}

\begin{document}
\author{G. Cordourier-Maruri\mbox{$^{1,2}$}}
\author{Y. Omar\mbox{$^{3}$}}
\author{R. de Coss\mbox{$^{1}$}}
\author{S. Bose\mbox{$^{2}$}}
\affiliation{
\mbox{$^{1}$}Departamento de F\'isica Aplicada, Cinvestav-M\'erida, A.P. 73 Cordemex, M\'erida, Yucat\'an 97310, M\'exico\\
\mbox{$^{2}$}Department of Physics and Astronomy, University College London, Gower Street, London WC1E 6BT, United Kingdom\\
\mbox{$^{3}$}Physics of Information Group, Instituto de Telecomunica\c{c}\~oes, P-1049-001 Lisbon, Portugal, CEMAPRE, ISEG, Universidade T\'{e}cnica de Lisboa, P-1200-781 Lisbon, Portugal}

\begin{abstract}

We show that the feature of Klein tunneling makes graphene a unique interface for implementing low control quantum gates between static and mobile qubits. A ballistic electron spin is considered as the mobile qubit, while the static qubit is the electronic spin of a quantum dot fixed in a
graphene nanoribbon. Scattering is the low control mechanism of the gate, which, in other systems, is really difficult to exploit because of both back-scattering and the momentum dependence of scattering. We find that Klein tunneling enables the implementation of quasi-deterministic quantum gates regardless of the momenta or the shape of the wave function of the incident electron. The
Dirac equation is used to describe the system in the one particle approximation with the  interaction between the static and the mobile spins modelled by a Heisenberg Hamiltonian.  Furthermore, we discuss an application of this model to generate entanglement between two well separated static qubits.

\end{abstract}


\title{Graphene Enabled Low-Control Quantum Gates between Static and Mobile Spins}
\maketitle
Interfacing static and mobile qubits is one of the central tasks in the emerging area of quantum technology, as it may aid in linking separated quantum registers. As in any other 
quantum information processing (QIP) task, it has to deal with the twin challenges of decoherence and  demanding control. The latter challenge can be somewhat alleviated in a scenario which 
uses the scattering of one flying qubit with a static qubit to accomplish an useful QIP scheme. In
this way we just have to set the initial configuration, inject the flying qubit and measure the final state. In a solid state scenario, the experimental
development of the new field called quantum electron optics \cite{hermeline,mcnell} allows the possibility to manipulate the path of just one
electron in a ballistic regime. Thus the flying qubit can be implemented with a ballistic electron spin. The static qubit can be created with the
spin of a quantum dot (QD) or a magnetic impurity, embedded on a quantum wire, which can be implemented with quantum Hall edge states 
\cite{hermeline,mcnell}, with carbon nanotubes \cite{and} or a graphene nanoribbon \cite{tra} (we adopt this last realization in this paper). Thus all basic ingredients for interfacing static and mobile spins through scattering already exist, even though they have not been put together in one setup. 

 However, as soon as one starts to consider scattering as a low control mechanism for interfacing static and mobile qubits, a number of obstacles seem to present. The most obvious is the spatial scattering in unwanted directions when the spins interact so that it is difficult to maintain the directionality of the mobile qubit. This is important if the mobile qubit has to interact with another static qubit in series and thereby connect two quantum registers. Despite this, a number of useful QIP applications of the scattering of a static and a mobile qubit have been presented \cite{costa,yuasa,ci1,ci3,hida,cor,ci5,gun} using techniques such as the post-selection of events were the mobile spin was spatially scattered in a given direction. Essentially this makes the protocols non-deterministic. The most useful task, namely accomplishing a unitary quantum gate between the static and the mobile spin through scattering is very challenging, requiring additional mirrors and/or potential barriers for the mobile particle, as well as a careful placement of these elements
\cite{cor,ci5}. Central to most schemes is the necessity to engineer the incident states of the mobile spins to a very narrow wavepacket in momentum space (nearly monochormatic electrons). Arbitrary wavepackets of the mobile spins would be a significant hindrance to such protocols. 
In this paper, we show that
graphene is a unique host material that can help to overcome the problems of both back-scattering and the momentum dependence of scattering while interfacing a static and a mobile spin qubit. In particular, we find that a graphene nano-ribbon can enable quasi-deterministic quantum gates between a ballistic electron spin and a quantum dot spin in which one does not need to finely control the incident wavepacket of the ballistic electron. While a graphene nano-ribbon is already appreciated as an ideal material for quantum spintronics by minimizing the decoherence problem \cite{tra,graphene-dec1}, our work indicates that it would also be act as an ideal interfacing element between a static and a mobile qubit.

Graphene is a monolayer of carbon atoms packed into a hexagonal crystal structure with extraordinary electronic properties 
\cite{castro} and presents a high spin coherence time. The last is due to the low spin-orbit coupling in a carbon-based materials, and because
natural carbon consists predominantly of zero-spin isotope $^{12}$C, for which the hyperfine interaction is absent \cite{tra,sil}. All this makes
graphene a very interesting option to implement spintronic systems. The main feature of the electronic properties in graphene is the linear dependence between energy and momentum of $2p_z$ electrons at low energies,
making a difference with the usual quadratic energy - momentum dispersion relation in ordinary materials. This peculiar relation causes the electron
transport in graphene to be governed by the Dirac-like Hamiltonian \cite{castro}, and electrons behave like massless Dirac fermions. The relativistic
analogy extends to the electron wavefunction in graphene, which is a two-component vector discriminating between the contribution of the two triangular
sublattices, constituting the hexagonal graphene lattice. This degree of freedom is known as {\it pseudospin} and its states have a well defined
chiriality: the pseudospin is parallel (antiparallel) to the electron (hole) momentum \cite{castro}. To perform a pseudospin flip process, an electron must
interact with a short range potential acting differently on the two sublattices. In the rest we consider that the pseudospin is conserved. Then, if a
ballistic graphene electron is scattered by a potential barrier in one direction, a hole state moving in the opposite direction is created inside the
potential, to preserve the pseudospin direction. This allows a perfect electronic transmission through a potential barrier by means of a hole state,
analogically to a relativistic transport phenomena known as Klein tunnelling \cite{castro,tra,kat,bab,young,chei}. 

Although the presence of Klein tunnelling increases coherence in the ballistic electrons, it also represents a limitation to create quantum confinement. 
The use of graphene nanoribbons with semiconducting armchair boundaries solves this problem; the transversal direction confinement in a nanoribbon with this conditions creates a gap between conduction and valence band \cite{bre,two,kat}. This energy gap can be use to localize electrons in a small region
through electric gating \cite{tra}. In this letter, we consider a graphene nanoribbon of width $W$ and with semiconducting armchair boundaries in the $y$
axis (see \fig 1). The transversal confinement produces a quantization of the transverse wavevector $k_y$, which can be expressed as \cite{tra,bre}

\begin{equation}\label{kyn}
k_{ny}=\left(n \pm \frac{1}{3}\right)\frac{\pi}{W} \ \ \ \ \ \ \ \ \ \ n \in\mathbb{Z} .
\end{equation}     

The band gap is $E_{gap}=2\hbar v_F k_{0y}$, with $v_F$ the graphene Fermi velocity ($v_F \approx 10^6$ m s$^{-1}$ \cite{kat}), which can be used to make a
one-electron QD with a square potential well \cite{tra}. Suppose that the QD has rectangular shape, with the same width as the graphene nanoribbon and
with a length $\Delta x$. Then, a ballistic electron moving along on the graphene nanoribbon is scattered by the QD, with incident angle $\theta$
and energy $\epsilon_b$, as is shown in \fig 1. The transversal confinement is a constraint to the incident angle $\theta$, as the quantization of 
$k_y$ limits the accessible wave vectors with a fixed energy $\epsilon_b$ as

\begin{equation}\label{theta}
\theta = \tan^{-1}\left[\frac{k_{ny}}{\sqrt{\epsilon_b^2/\hbar^2v_F^2-k_{ny}^2}}\right].
\end{equation} 

\noindent
The problem can be overcome tuning the ballistic electron energy $\epsilon_b$. The frontal scattering ($\theta = 0$) is asymptotically obtained when
$\epsilon_b \gg \hbar v_F k_{ny}$. This is a constraint in the use of Klein tunnelling in semiconducting armchair graphene nanoribbons, as the total
transmission is only present in frontal scattering \cite{kat,comm}. Nevertheless, in the next we will show how variations in the incident angle produce
only small changes in the transmission rates.   

Consider that the ballistic electron energy $\epsilon_b$ and $\Delta x$ are such that we can assume $5\Delta x < \lambda$, where 
$\lambda = 2\pi \hbar v_F/\epsilon_b$ is the ballistic electron de Broglie wavelength, in order to avoid resonant behaviours. This assumption allows us to
describe the ballistic electron and QD interaction as a delta potential. For example, with a ballistic electron energy $\epsilon_b = 60$ meV, the Fermi
wavelength is $\lambda \approx 100$ nm, then $\Delta x$ could be $21$ nm. Although the Coulomb interaction between the QD and the incident electron is
considered as long range, this range can be controlled with the addition of a metallic substrate, which screen the QD charge \cite{jun}. Then, the system can
be depicted with a one-particle Dirac-like Hamiltonian 

\begin{equation}\label{hgra}
\hat{H} =-i\hbar v_F \mathbf{\sigma}\cdot\nabla+ J\hat{\mbox{\boldmath$S$}}_e\cdot\hat{\mbox{\boldmath$S$}}_i\delta(x),
\end{equation} 

\noindent
where $\mathbf{\sigma}=\{ \sigma _x,\sigma_y\}$ is the Pauli matrices vector, and the QD is located at $x=0$. Note that here we have assumed that the scattering process is
elastic, then the wavevector of the incoming and the outcoming electron are the same. This is indeed well within the parameter scale of our problem as theoretical studies have estimated the inelastic length and time scales for hot ballistic electrons in graphene to be $100$ nm and $0.1$ps respectively \cite{das-sarma}. The elasticity also guarantees that the configuration -- one electron confined to the dot, and one electron moving, remains fixed even after the scattering. Treating the quantum dot spin in a manner similar to an Anderson impurity, the effective Heisenberg term $J\hat{\mbox{\boldmath$S$}}_e\cdot\hat{\mbox{\boldmath$S$}}_i\delta(x)$ in Eq.(\ref{hgra}) can be found out to reduce the two electron problem to a one electron problem. Note that because of the delta function, the units of $J$ is eV\AA, rather than just eV.

\begin{figure}[h] \label{figura2}
     \begin{center}
     \resizebox{9cm}{!}{\includegraphics{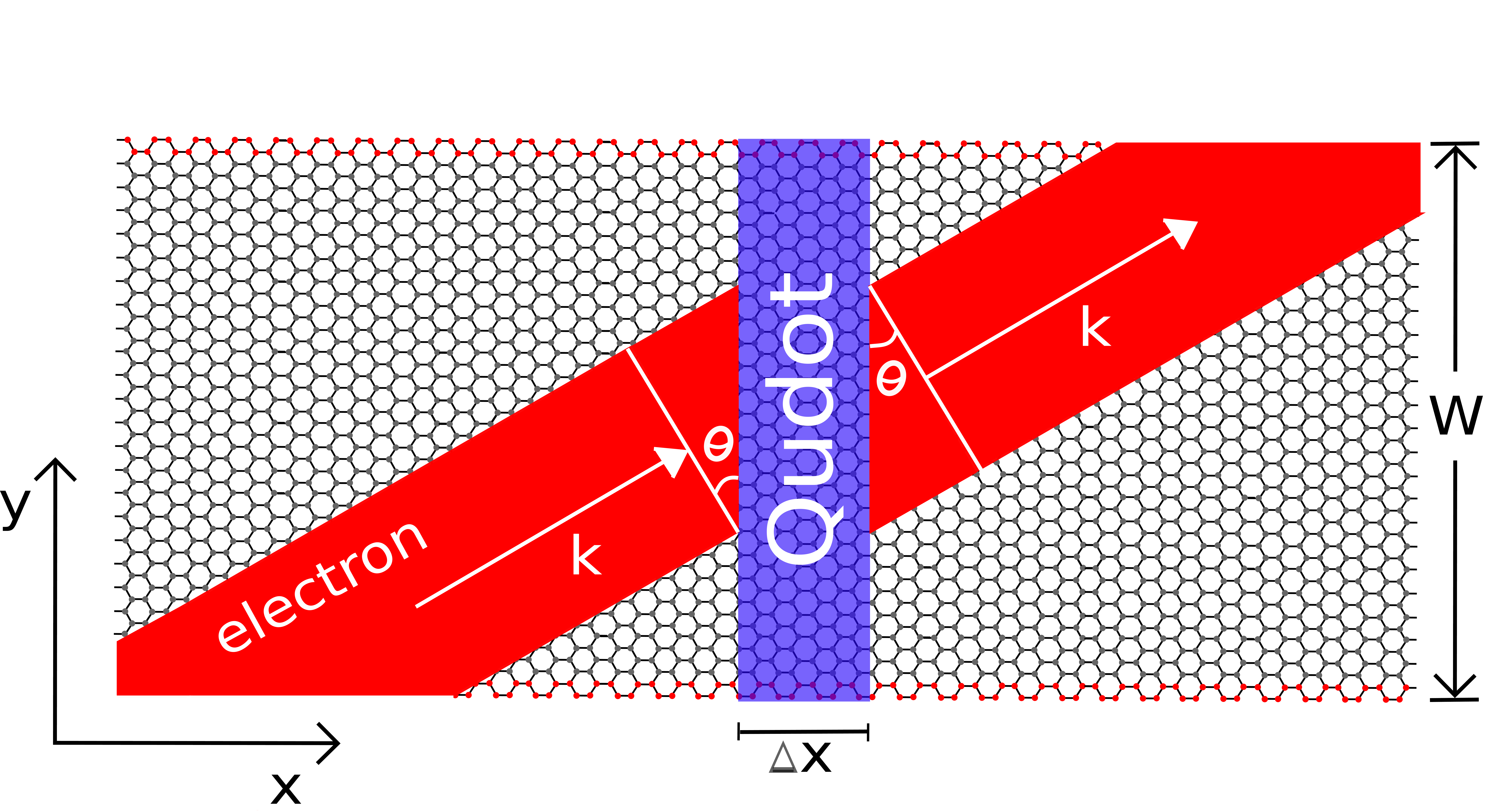}}
     \caption{Schematic diagram of an electron elastic scattering with an incident angle $\theta$, on a rectangular QD of $\Delta x$ lenght.}
     \end{center}
\end{figure}
   
The eigenstates of the Hamiltonian (\eq\ref{hgra}) must include the pseudospin, electron and QD spins contributions. We consider the QD as a static 
$1/2$ - spin particle, with the wavefunction

\begin{equation}\label{wfi}
\psi_d=A\chi_{1/2} + B\chi_{-1/2} = A\left( \begin{array}{c} 1 \\ 0 \end{array}\right)+ B\left( \begin{array}{c} 0 \\ 1 \end{array}\right),
\end{equation}

\noindent
in terms of the Pauli vectors $\chi_{\pm 1/2}$. The electron wavefunction in a graphene nanoribbon can be modelled as a two-dimensional monochromatic
wave moving with a wavevector $k = \epsilon_b / \hbar v_F$ with $k^2= k_x^2+k_y^2$. Then the ballistic electron wavefunction in the region $x<0$ is a 
four-component vector (two from the pseudospin and two from the spin) of the form 

\begin{eqnarray}\nonumber\label{wfe1}
\psi_e & = & \left( \begin{array}{c} 1 \\ s e^{i\theta} \end{array}\right)\otimes\left(a\chi_{1/2} + b\chi_{-1/2}\right)e^{i(k_xx+k_yy)} +
 \\
& & \left( \begin{array}{c} 1 \\ -s e^{-i\theta} \end{array}\right)\otimes\left(c\chi_{1/2} + d\chi_{-1/2}\right)re^{i(-k_xx+k_yy)}.
\end{eqnarray}

\noindent
The first term in the right side of \eq\ref{wfe1} represents the incident electron while the second term describes the reflected electron with a probability
amplitude $r$. The transmitted electron wavefunction has the form 

\begin{equation}\label{wfe}
\psi_e = \left( \begin{array}{c} 1 \\ s e^{i\theta} \end{array}\right)\otimes\left(f\chi_{1/2} + g\chi_{-1/2}\right)te^{i(k_xx+k_yy)},
\end{equation}

\noindent
for $x>0$ with a transmission probability amplitude $t$. Here, $s$ = sgn$(\epsilon_b)$ generalizes the wavefunction for the case of a hole, considered as a 
negative-energy particle. The incident angle $\theta = \tan^{-1}(k_y/k_x)$ is measured from the $x$ axis, and it also depicts the pseudospin direction. In
the one-particle approximation the entire system wavefunction can be written as $\Psi = \psi_e \otimes \psi_d$. $\Psi$ is a eight-dimension vector with
twelve probability amplitudes in the two regions ($x<0$ and $x>0$). Similarly to the well known nonrelativistic delta barrier problem, to find the boundary
conditions we have to integrate the \eq\ref{hgra} on an infinitesimally small interval around $x=0$ as 

\begin{equation}\label{bo1}
\lim_{\Delta x \to 0} \int_{-\Delta x}^{\Delta x}
\hat{H}\Psi(x,y)dx =\lim_{\Delta x \to 0} \int_{-\Delta x}^{\Delta x} \epsilon_b \Psi(x,y)dx.
\end{equation}

\noindent
A special problem for this limit evaluation is that in a Dirac-like Hamiltonian the inclusion of a delta potential produces a discontinuity in the
wavefunction. We will use the approach of \cite{sub} to solve this problem, and
we allow the components of $\Psi(x,y)$ to have a finite discontinuity at $x=0$ and extend the definition of delta function by writing 

\begin{equation}\label{delta}
\lim_{\Delta x \to 0} \int_{-\Delta x}^{\Delta x} \psi(x,y)\delta(x)dx = \frac{\psi(0_+,y)+\psi(0_-,y)}{2}.
\end{equation}

After applying \eq\ref{delta} in \eq\ref{bo1}, we arrive to the boundary condition

\begin{eqnarray}\nonumber\label{bo2}
0 & = & -\imath\hbar v_F\mathbf{\sigma}_x[\Psi(0_+,y)-\Psi(0_-,y)] +
 \\
& & \frac{J}{2}\hat{\mbox{\boldmath$S$}}_e\cdot\hat{\mbox{\boldmath$S$}}_i[\Psi(0_+,y)-\Psi(0_-,y)].
\end{eqnarray}

\noindent
Assuming we know the initial spinor state of the electron and quantum dot, we can solve the 12-variable equation system with the initial conditions and 
\eq\ref{bo2}. Now we focus on the probability amplitude of the scattered wavefunction if a spin flip takes place ($t_s$), or if it does not ($t_n$), which 
are

\begin{equation}\label{tn}
t_n =\frac{64v^2_F\hbar^2-3J^2}{64v^2_F\hbar^2-16iv_F\hbar J+3J^2} + O(\theta^2),
\end{equation}

\noindent
and 

\begin{equation}\label{ts}
t_s =\frac{32iv_F\hbar J}{64v^2_F\hbar^2-16iv_F\hbar J+3J^2} + O(\theta^2).
\end{equation} 

\noindent

  We now pause breifly to state the non-trivial conditions needed for a unitary quantum gate between a static and a scattered qubit dependent on the latter's transmission \cite{cor}. Usually, only the whole scattering process is unitary in the space and spin combined degree of freedom. If spin-flipped and no spin-flipped states of the mobile spin have different probablities for transmission, then, by measuring the transmission, some information about the spin state can be acquired. This transformation is then clearly not unitary in the spin degree of freedom.
The general transformation acting on the spin density matrix $\rho$ of the two qubits on transmission of the mobile spin is
\begin{equation} \label{nonlinear}
\rho' = \frac{\mathbf{T}\rho \mathbf{T}^{\dagger}}{\mathrm{Tr}\,[\mathbf{T}\rho\mathbf{T}^{\dagger}]},
\end{equation} 
\noindent
where $\mathbf{T}$ is the probability amplitude matrix acting in a nonlinear way on $\rho$. The condition needed to assure unitarity and linearity of 
$\mathbf{T}$ is $|t_n + t_s| = |t_n - t_s|$, with $t_s$ and $t_n$ the probability amplitudes of do or do not have a spin flip of both spins after scattering,
respectively. This is the condition to implement an electron scattering quantum gate \cite{cor}. Not only is it very intricate to satify $|t_n + t_s| = |t_n - t_s|$ for particles with Schroedinger dispersion, the total transmission $|t_n|^2+|t_s|^2$ in such cases is significantly lower than unity for really useful two qubit gates, making the gates non-deterministic \cite{cor}. 

Now notice from Eqs.(\ref{tn}) and (\ref{ts}) that the Klein tunnelling ($|t_n|^2+|t_s|^2=1$) is present, and the gate condition $|t_n + t_s|=|t_n - t_s|$ is fulfilled independently of $J$ when 
$\theta=0$. Then, the gates implemented will be deterministic. The cause of this behaviour can be seen clearly if we express the system spinor in terms of
the singlet-triplets basis ($\{\psi_-,\ket{\uparrow\uparrow},\psi_+,\ket{\downarrow\downarrow}\}$), which are eigenfunctions of the Heisenberg operator,
then the dynamics of the singlet and triplet subspaces are decoupled. In each of these subspaces, the Heisenberg interaction term of \eq\ref{hgra} reduces
to a spinless potential barrier \cite{cor} so the effective Hamiltonian describes a particle scattering from two spin-independent potentials as

\begin{equation} \label{Heff}
\hat{H}_{S}=-i\hbar v_F \mathbf{\sigma}\cdot\nabla +V_S\,\delta(x),
\end{equation} 

\noindent
where 

\begin{equation} \label{Vs}
V_S=\frac{J}{2} [S (S+1)-3/2]\,\,
\end{equation} 

\noindent
is an effective potential and $S$ is the quantum number associated with $\hat{\mathbf{S}}^2$ and $S\!=\!0$ ($S\!=\!1$) in the case of the singlet 
(triplet). The ballistic electron, as a massless pseudo Dirac fermion, can be perfectly transmitted through these potentials due to the Klein tunnelling
\cite{tra,castro,kat,bab,young}, and we expect that $|t_n + t_s|=|t_n-t_s|$ for any value of $J$. In \eqs\ref{tn} and \ref{ts} we can see that the
angular dependence of the probability amplitudes in frontal insertion is resilient to small angular changes. Also, in frontal scattering the Klein
tunnelling is independent of $k$, and then it is present when the ballistic electron is in a wavepacket form. In \fig 2 a) we show the evolution of the
probabilities of detecting a transmitted electron, i.e.\ the success probability of the gate, $|t_n|^2 + |t_s|^2$ as a function of the incident angle $\theta$,
and the coupling factor $J$. Notice that the values for the success of the gate change only approximately 5 \% when $\theta$ changes from $0$ to $\pm \pi/16$,
and remains almost constant as $J$ evolves. 

The strength of the Heisenberg spins interaction is described by the coupling factor $J\approx4\Delta x t^2/U $, which contains all the information about the
shape of the quantum dot through the overlap integral

\begin{equation}\label{ove}
t = \epsilon_b\int\psi_b(\pmb{r})\psi_d(\pmb{r})dr, 
\end{equation} 

\noindent
and the Coulombic repulsion between electrons   

\begin{equation} \label{U}
U = \frac{e^2}{4 \pi \epsilon_0}\int\int\frac{|\psi_d(\pmb{r_1})|^2|\psi_b(\pmb{r_2})|^2}{|\pmb{r_1}-\pmb{r_2}|+\delta}dr_1dr_2, 
\end{equation} 

\noindent
where $\delta=0.0814$ nm is the radial extent of a $\pi$-orbital in graphene \cite{gun}, and $\psi_b$ and $\psi_d$ are the ballistic electron and QD
wavefunctions, respectively. Here, the exchange interaction between QD and ballistic electron depends on the parameters $w$, $\Delta x$ and the QD and
ballistic electron energies, which can be controlled to reach a desired $J$ value.

\begin{figure}[h] \label{figura3}
     \begin{center}
     \resizebox{8cm}{!}{\includegraphics{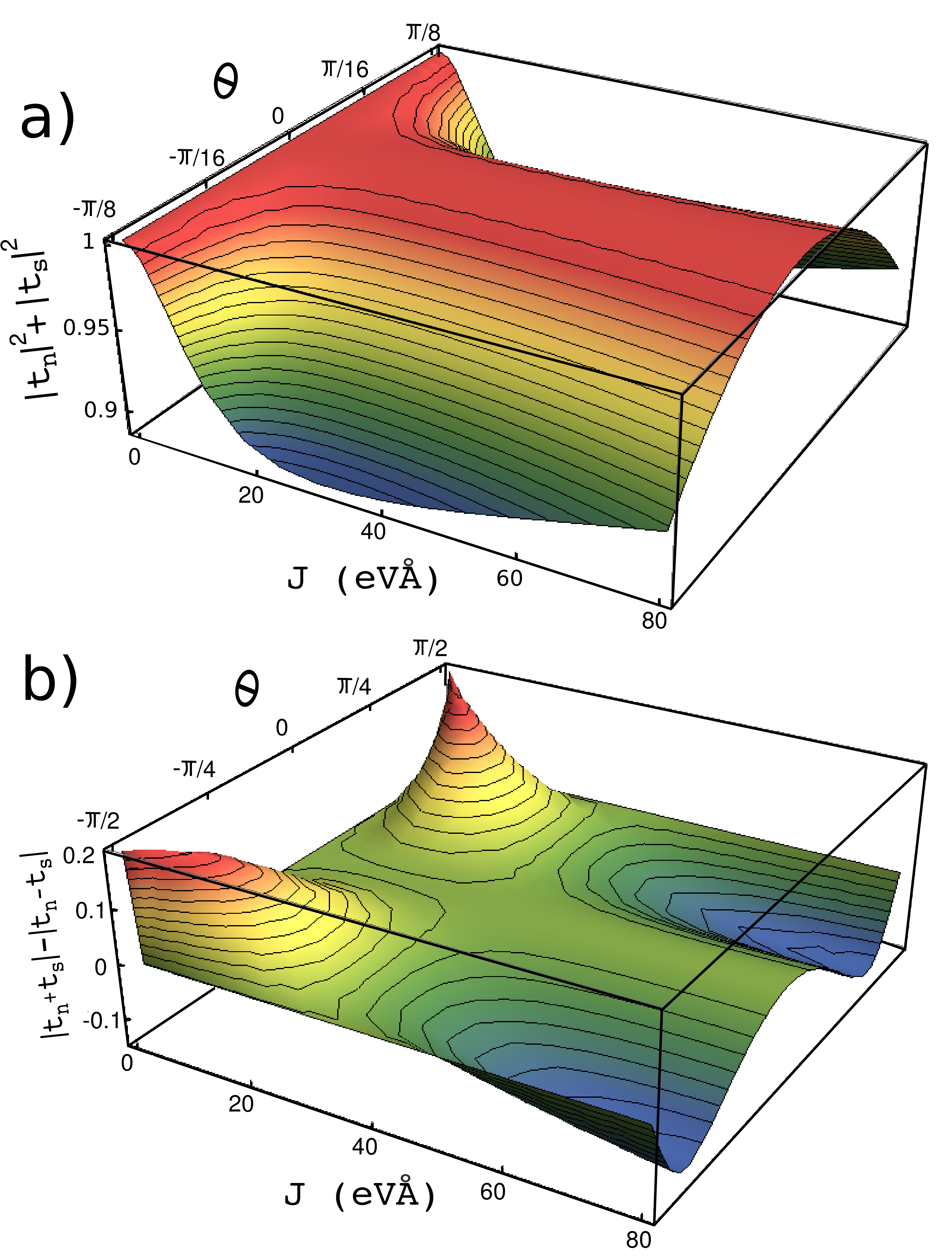}}
     \caption{Evolution of a) the gate probability of success ($|t_n|^2 + |t_s|^2)$ and b) the gate condition $|t_n + t_s|-|t_n - t_s|$ as a function of the
              electron incident angle ($\theta$) and the coupling factor ($J$).}
     \end{center}
\end{figure}

We show $|t_n + t_s|-|t_n - t_s|$ as a function of variations in  $\theta$ and $J$ in the \fig 2 b), whenever $|t_n + t_s|-|t_n - t_s| = 0$ the gate
condition is fulfilled. Notice that this is satisfied in $J=8\sqrt{1/3}\hbar v_F \approx 30$ eV\AA, independently of the angle of incidence. The gates
implemented here will be of the $\mbox{\emph{SWAP}}$ type and, with the exception of the one at $\theta=0$, they will be nondeterministic.

\begin{figure}[h]\label{figura4}
     \begin{center}
     \resizebox{8cm}{!}{\includegraphics{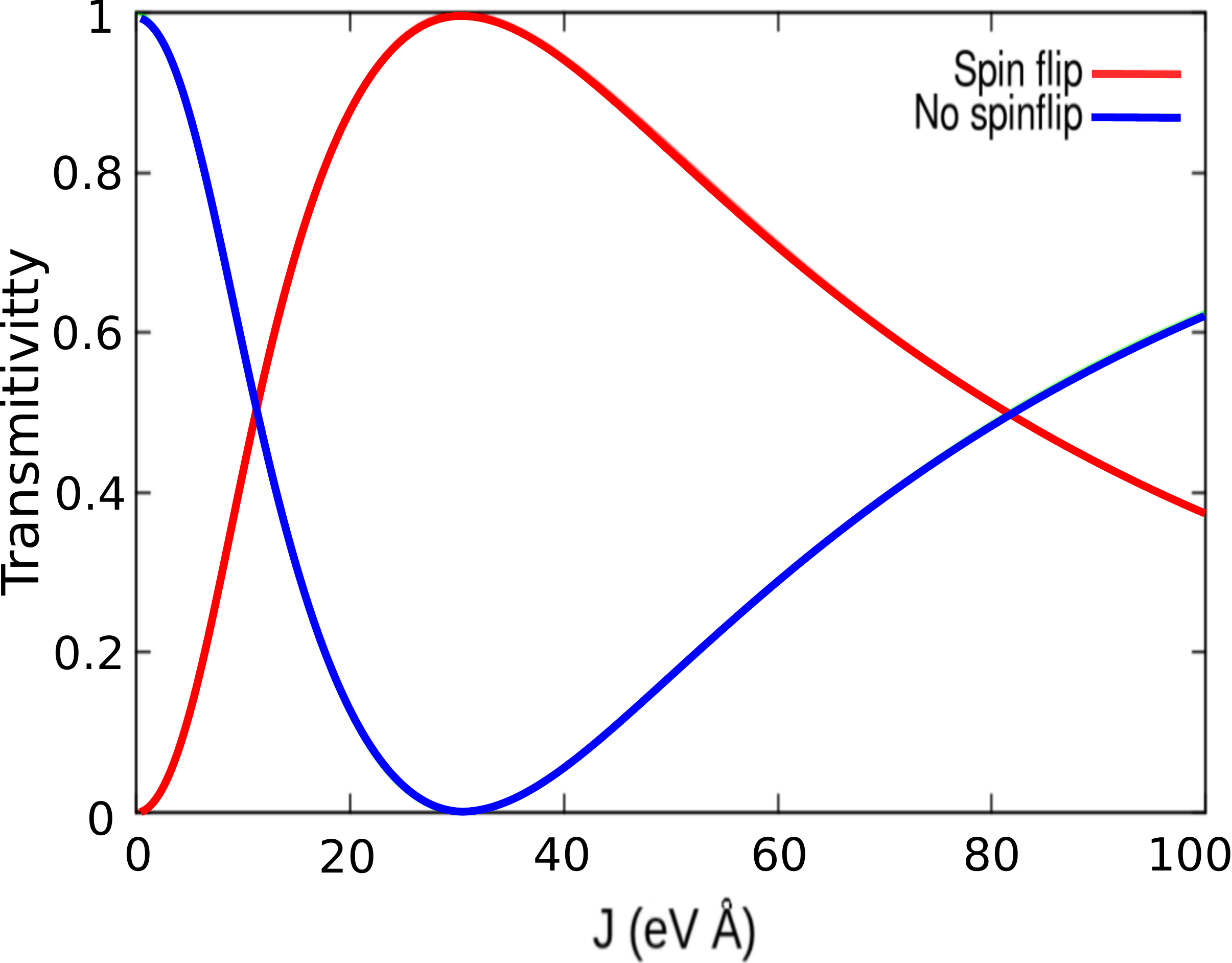}}
     \caption{Transmitivitty or probability of no spin flip ($|t_n|^2$ blue line) and of spin flip ($|t_s|^2$ red line) after a frontal scattering, as a function of the coupling factor $J$.}
     \end{center}
\end{figure}

In \fig 3 we show the transmitivitty or probability of the occurence of no spin flip ($|t_n|^2$ blue line) and of a spin flip ($|t_s|^2$ red line) after a frontal scattering, as a function of the coupling factor $J$. Notice there are two values of $J$ (namely, 
$J=8\hbar v_F\sqrt{11 - 4 \sqrt{7}}/3 \approx 11.2$ eV\AA \ \ and $J=8\hbar v_F\sqrt{11 + 4 \sqrt{7}}/3 \approx 81$ eV\AA) for which $|t_n|^2 =|t_s|^2$. At
those points we expect that the implemented gate will be a $\sqrt{\mbox{\emph{SWAP}}}$, which generates maximum entanglement between the electron and QD
spins. We see also a perfect spin flip point ($J \approx 30$ eV\AA), and the implemented gate will be a $\mbox{\emph{SWAP}}$. The implementation of this kind
of gates can be useful to initialize and to read out the QD spin state injecting a polarized ballistic electron whose final state can be measured directly.

A possible application of the previous model is to use the ballistic electron spin as a mediator to correlate two fixed and distant QD. The separation between
the two QD can be in the $100$ nm range because essentially there is no inelastic scattering over this range \cite{das-sarma}. Now suppose that we control the initial spin conditions and
the values of $J$ in each QD. In this way, we can set the probability amplitudes of the electron frontal scattering with the first QD to be $t_{n1}$ and 
$t_{s1}$, and also we can set the probability amplitudes of the electron frontal scattering with the second QD to be $t_{n2}$ and $t_{s2}$. If we inject a
ballistic electron with a known spin state, for instance $\ket{\uparrow}_e$, the transformation can be represented in the computation basis of the two QDs
spins ($\ket{\uparrow}_1\ket{\uparrow}_2$, $\ket{\uparrow}_1\ket{\downarrow}_2$, $\ket{\downarrow}_1\ket{\uparrow}_2$, $\ket{\downarrow}_1\ket{\downarrow}_2$)
as  

\begin{equation}   \label{Tdos}
 \mathbf{T}_2 =\left( \begin{array}{cccc}
            1          &             t_{s2}              &              t_{s1}t_{n2}            &    0               \\
            0          &             t_{n2}              &              t_{s1}t_{s2}            &    t_{s1}          \\
            0          &               0                 &                 t_{n1}               &    t_{n1}t_{s2}    \\
            0          &               0                 &                    0                 &    t_{n1}t_{n2}     \end{array} \right).
\end{equation} 

\noindent
The triangular form of this matrix is due to the total transmission in the scattering events, making that the result on the second QD has no effect 
on the first one and no resonant behaviour is expected. If we set the $J$ factor of the second QD to be SWAP-like, which means $|t_{s2}|=1$ and $|t_{n2}|=0$,
it is straightforward to see that after the electron frontal scattering with both QDs, we will obtain a superposition in the QDs spin states of the form
 $\Psi_{12} = t_{s1}\ket{\uparrow}_1\ket{\downarrow}_2 + t_{s2}\ket{\downarrow}_1\ket{\uparrow}_2$. Thus, with this process we can control the level of
entanglement generated after the scattering. If we choose the $J$ coefficient of the first QD to generate a $\sqrt{\mbox{\emph{SWAP}}}$ gate, which this time
means $|t_{s1}|=1/\sqrt{2}$ and $|t_{n1}|=1/\sqrt{2}$, we can assure a total entangled final state between QD spins. It is not necessary to perform a
postselection of any kind on the electron after the frontal scattering. 

 Before concluding, let us point out that one could possibly choose different parameter regimes to obtain the values of $J$ required above for the $\sqrt{\mbox{\emph{SWAP}}}$ and $\mbox{\emph{SWAP}}$, which are really the useful for QIP. However, for a parameter regime that we could readily identify, namely, $W=30$, $\epsilon_b = 60$ meV, $\Delta x = 21$ nm, one obtains an angle of incidence $\sim 22$ degrees ($90$\% reflection occurs at 20 degrees), $U= 78$ meV, $t=20$ meV, and thereby $J \sim 4.1$eV\AA. For this value the spin-flip probability on transmission is $\sim 0.1$, and thereby the entangled state of the impurity 1 and the ballistic spin that is generated is $\sim 0.95\ket{\uparrow}_e\ket{\downarrow}_1+0.31 \ket{\downarrow}_e\ket{\uparrow}_1$. This is still an entangled state, and the fact that this is a result of a unitary operation makes the gate an entangling gate -- which is still very useful for QIP, though not as readily useful as the maximally entangling $\sqrt{\mbox{\emph{SWAP}}}$ gate. For example, taking the scenario of two static quantum dot based spins $1$ and $2$ mentioned above, but with the ballistic electron, initially in the $\ket{\uparrow}_e$ state, undergoing the same gate with both spins $1$ and $2$ (as opposed to the above case) in the parameter set of this paragraph, one can create the highly entangled state $0.69 \ket{\downarrow}_1 \ket{\uparrow}_2+0.73 \ket{\uparrow}_1 \ket{\downarrow}_2$ with a $0.18$ success probability conditional on detecting the ballistic electron in the state $\ket{\downarrow}_e$ (say, by a spin filter).

In summary, in this work we have shown how the Klein tunnelling, present in the graphene electrons scattering off a rectangular QD, is useful to implement a
quasi-deterministic two-qubit quantum gate between the ballistic electron spin and the QD spin. The transversal confinement limits the incident angle in 
the scattering process, due to the quantization of the transverse wavevector $k_y$. This problem can be overcome tuning the ballistic electron energy to reach the frontal scattering ($\theta=0$) angle asymptotically. We show that when $J=8\sqrt{1/3}\hbar v_F \approx 30$ eV\AA, $|t_n|^2$ is equal to zero for any value of the incident angle, and a $\mbox{\emph{SWAP}}$ gate is
obtained. In a frontal scattering, the Klein tunnelling is present and we always will find quantum gates. The gates implemented in these conditions are
quasi-deterministic, because the gate success depends on how approximately frontal the scattering is. However, we see that a change in the incident angle
of $\pm \pi/16$ from the ideal frontal angle produce only small changes (of about 5 \%) in the gate success probability. It has also been shown how this model
can be used to generate and control the entanglement between two fixed and distant magnetic impurities, taking a ballistic electron spin as a mediator. Even in a parameter set that we could readily identify, an useful entangling quantum gate is obtained between the static spin and the ballistic electron spin. Perhaps some of this work will be adaptable to a setting of carbon nanotubes, were an absence of back-scattering is also present \cite{and}. 
        
\section{Acknowledgments} \label{ack}

YO thanks the support from project IT-PQuantum, as well as from Funda\c{c}\~{a}o para a Ci\^{e}ncia e a
Tecnologia (Portugal), namely through programme POC\-TI/PO\-CI/PT\-DC and projects PEst-OE/EGE/UI0491/2013, PEst-OE/EEI/LA0008/2013 and 
PTDC/EEA-TEL/103402/2008 QuantPrivTel, partially funded by FEDER (EU), and from the EU FP7 projects LANDAUER (318287) and PAPETS (323901). SB is supported by the ERC. This research was partially funded by the Consejo Nacional de Ciencia y Tecnolog\'ia (Conacyt, M\'exico) under grant No. 46630-F. We thank Guido Burkard, Mark Buitelaar and John Jefferson for clarifying some aspects of the background literature.


\begin{thebibliography}{16}
\bibitem{comm} There are another angles of total transmission due to Klein tunneling, but they depend of the resonant conditions in the geometry
                 of the system, which are only valid for monochromatic ballistic electrons.
\end{thebibliography}
\end{document}